\documentclass[12pt]{article}
\usepackage{graphicx}
\newcommand{\preprint}[1]{\begin{flushright}#1\end{flushright}}

\newcommand \bra[1]{\left< {#1} \,\right\vert}
\newcommand \ket[1]{\left\vert\, {#1} \, \right>}
\newcommand \braket[2]{\hbox{$\left< {#1} \,\vrule\, {#2} \right>$}}
\newcommand{\bea}{\begin{eqnarray}}
\newcommand{\eea}{\end{eqnarray}}
\newcommand{\simgt}{\hbox{ \raise3pt\hbox to 0pt{$>$}\raise-3pt\hbox{$\sim$} }}
\newcommand{\simlt}{\hbox{ \raise3pt\hbox to 0pt{$<$}\raise-3pt\hbox{$\sim$} }}
\newcommand \vc[1]{{\bf {#1}}}

\def\to{\rightarrow}

\begin{document}
\preprint{hep-ph/9903498\\TU-562\\March 1999}
\vspace*{3cm}
\begin{center}
  {\bf\large
    ${\cal O} ( \alpha_s^2 )$ Corrections to $e^+e^- \to t\bar{t}$
    Total and Differential \\
    Cross Sections Near Threshold
    }
  \\[5mm]
  {
    T. Nagano, A. Ota and Y. Sumino
    }
  \\[5mm]
  {\it
    Department of Physics, Tohoku University\\
    Sendai, 980-8578 Japan
    }
\end{center}
\vspace{3cm}
\begin{abstract}
Recently full ${\cal O}(\alpha_s^2,\alpha_s\beta,\beta^2)$
corrections to the threshold total cross section
for $e^+e^- \to t\bar{t}$ have been calculated, and the 
reported corrections turned out to be unexpectedly large.
We study how to reduce theoretical uncertainties
of the cross section.
We adopt a new mass definition proposed by Beneke, which
incorporates a renormalon-pole cancellation in the total energy
of a static quark-antiquark pair. 
This improves the convergence of the $1S$ resonance mass, while
the normalization of the cross section scarcely changes.
We argue that resummations of logarithms are indispensable, 
since two 
largely separated scales dictate the shape of the cross section.
As a first step, we resum logarithms in the Coulombic part of the
$t\bar{t}$ potential and observe a considerable improvement in the
convergence of corresponding corrections.
There still remain, however, large corrections,
which arise from a $1/r^2$ term in the $t\bar{t}$ potential.
We also calculate full
${\cal O}(\alpha_s^2 , \alpha_s \beta , \beta^2)$ corrections
to the momentum distributions of top quarks in the
threshold region.
Corrections to the distribution shape are of moderate size
over the whole threshold region.
\end{abstract}

\newpage
\section{Introduction}
\label{s1}

The top quark pair production in the threshold region
at future $e^+e^-$ or $\mu^+\mu^-$ colliders 
is considered as an ideal process for precision measurements of
top quark properties.
Already many works have been devoted to the analyses
of this process both theoretically and experimentally
\cite{fk}-\cite{kt}.

Recently full ${\cal O}(\alpha_s^2 , \alpha_s \beta , \beta^2)$ corrections 
to the total cross section for
$e^+e^- \to \gamma^* \to t\bar{t}$ in the threshold region
have been calculated 
independently by \cite{ht,mye} using the NRQCD formalism.\footnote{
Corrections induced by the
axial-vector coupling to a $Z$-exchange have been calculated,
which also contribute as  
${\cal O}(\alpha_s^2 , \alpha_s \beta , \beta^2)$ corrections
\cite{fk2,kt}.
}
Both calculations showed that these corrections are surprisingly large.
Moreover, they found very poor convergence 
of the cross section as they compared
the leading-order (LO), next-to-leading order (NLO) and
next-to-next-to-leading order (NNLO) calculations.
Theoretically, the calculation in \cite{mye} is more sophisticated in that 
in the vicinity of each resonance pole it includes all
${\cal O}(\alpha_s^2)$ corrections to the resonance mass
and to the residue.
(Practically, 
the location of the $1S$ resonance peak will
provide an important information related to the top quark
mass.)
The two calculations were reproduced in \cite{yakovlev}, where
some numerical error of \cite{mye} was corrected.
There appeared other observations which noted potentially large
theoretical uncertainties on different grounds \cite{jkpst,yy}.

In this paper, we first study how to cure the problem of the bad
convergence of the total cross section observed in the above works.
One possible modification is to redefine the top quark mass.
It was found \cite{hssw,beneke,private} that a renormalon pole
contained in the QCD potential between a static quark-antiquark pair
gets cancelled in the total energy of the pair
$2m_{\rm pole} + V_{\rm QCD}(r)$
if the pole mass $m_{\rm pole}$ is expressed in terms of
the $\overline{\rm MS}$ mass.
As a result, the series expansion of this total energy in 
the $\overline{\rm MS}$ coupling $\alpha_s(\mu)$ behaves
better if we use the $\overline{\rm MS}$ mass instead of
the pole mass.
This suggests that the $\overline{\rm MS}$ mass has a more natural
relation to
physical quantities of a static (or non-relativistic) quark-antiquark
system.
Beneke proposed a new quark mass definition, which incorporates a
renormalon pole cancellation, and which is
related to the $\overline{\rm MS}$ mass in a well-behaved
series \cite{beneke}.\footnote{
A problem is that the relation between the $\overline{\rm MS}$ mass and
the pole mass is known only up to ${\cal O}(\alpha_s^2)$
\cite{gbgs}.
Meanwhile, if we want to use the $\overline{\rm MS}$ mass in the
NNLO analyses of the threshold cross sections,
we need to know this relation up to ${\cal O}(\alpha_s^4)$, 
since the binding energies of the boundstates are 
$\sim \alpha_s^2 m$ already at LO.
}
We adopt this new mass definition and study the convergence 
properties of the $t\bar{t}$ threshold cross section.

As another improvement, we incorporate a log resummation in the
cross section.
There is a logical necessity for resummations of logarithms in
calculations of the total cross section in the threshold region.
This feature is qualitatively different from energy regions
far above the threshold.
In the vicinity of distinct resonance peaks (for a realistic top quark
this corresponds only to the $1S$ peak), the total cross section
takes a form
\bea
\sigma_{\rm tot}(s) \sim - \mbox{Im}
\sum_n \frac{|\psi_n(0)|^2}{\sqrt{s}-M_n+i\Gamma_n} .
\eea
The resonance spectra $M_n$'s are dictated by the shape of the
quark-antiquark QCD potential at the scale of Bohr 
radius $r \sim (\alpha_s m_q)^{-1}$,
while the wave functions at the origin $\psi_n(0)$'s are determined by
the shape of the potential at a considerably shorter distance, 
$1/m_q < r \ll (\alpha_s m_q)^{-1}$.
Thus, in order to predict reliably both the
energy dependence and normalization of
the total cross section in the resonance region, one needs to 
calculate the shapes of the QCD potential accurately at largely
separated two scales.
This naturally requires log resummations using renormalization-group
equations.
At NLO, a log resummation was incorporated first in \cite{sp}.
As a first step at NNLO, we resum logarithms in the 
Coulombic part of the $t\bar{t}$ potential in this work.

The second subject of this paper is a calculation of
full ${\cal O}(\alpha_s^2 , \alpha_s \beta , \beta^2)$ 
corrections to the momentum distribution of top quarks in 
the threshold region.
It is expected that the top momentum distribution
will provide important informations independent of 
those from the total cross
section \cite{sfhmn,jkt,jt,fms}.
We therefore study how the distribution are affected by the corrections.
We find that the 
sizes of corrections to the distribution shape are moderate in
comparison with the corrections to the total cross section.

We note here that in our analyses no consistent treatment of the decay
process of top quarks is attempted.
Following \cite{ht,mye} we 
merely replace the non-relativistic
Hamiltonian as
\bea
H_{\rm NR} \to H_{\rm NR} - i\Gamma_t ,
~~~~~~~~~~~~~~
(\Gamma_t:~\mbox{top-quark on-shell width})
\label{repl}
\eea
which is the correct prescription for calculating the total
cross section at LO \cite{fk} and 
at NLO \cite{my,thesis,cracow,fkm,fms}
(provided 
we include ${\cal O}(\alpha_s)$ corrections to $\Gamma_t$
\cite{r42,r53} at NLO).
At NNLO, corrections related to the top decay process 
have not been calculated yet.
As for the differential cross sections, the above prescription is
valid only at LO.
At NLO, the final-state interactions 
affect the differential
cross sections non-trivially in the
threshold region but cancel out in the total cross section 
\cite{thesis,cracow,fms,hjkp,ps}; see also \cite{schmidt,siopsis,bbc}.

In Section~\ref{s2} we recalculate the total cross sections at
LO, NLO and NNLO.
Then we incorporate a new mass definition in Section~\ref{s3}.
We examine the effect of a log resummation in the Coulombic potential
in Section~\ref{s4}.
The momentum distributions of top quarks including 
full ${\cal O}(\alpha_s^2)$ corrections are presented in Section~\ref{s5}.
Section~\ref{s6} contains summary and discussion.
In Appendix~\ref{appa} all notations and definitions are collected.
A derivation of the momentum distribution at NNLO is presented in
Appendix~\ref{appextra},
while in Appendix~\ref{appb} we prove the unitarity relation between
the total cross section and the momentum distribution.

\section{Total Cross Section}
\label{s2}

As derived in \cite{mye}, the photon-exchange contribution to 
the $e^+e^- \to t\bar{t}$ threshold total cross section including
full ${\cal O}(\alpha_s^2 , \alpha_s \beta , \beta^2)$ 
corrections is given by
\bea
\sigma_{\rm tot}(s) &=& 
\frac{32\pi^2 \alpha^2}{s^2} \, N_c Q_t^2
\left\{
1 
+ \left( \frac{\alpha_s(m_t)}{\pi} \right) C_F C_1
+ \left( \frac{\alpha_s(m_t)}{\pi} \right)^2 C_F C_2(r_0)
\right\}
\nonumber \\ &&
\times
{\rm Im} 
\left[
\left( 1 + \frac{E+i\Gamma_t}{6m_t} \right)
G(r_0,r_0)
\right] .
\label{totcs}
\eea
Here, $C_1$ and $C_2(r_0)$ are vertex renormalization constants;
their explicit forms are given in Appendix \ref{appa}.
The Green function is defined by
\bea
\left\{
-\frac{1}{m_t} 
\left[ \frac{d^2}{dr^2}+\frac{2}{r}\frac{d}{dr} \right]
+ V(r) - \left[ \, \omega + \frac{\omega^2}{4m_t} \right]
\right\} G(r,r') = \frac{1}{4\pi r r'} \, \delta (r-r') ,
\label{green}
\eea
where
\bea
&&
V(r) = V_{\rm C}(r) - 
\frac{3 \omega }{2 m_t}\frac{C_F \alpha_s(\mu)}{r}
- \frac{C_F (3C_A+2C_F)\alpha_s(\mu)^2}{6 m_t r^2} 
,
\\ &&
V_{\rm C}(r) = - \, C_F \frac{\alpha_s(\mu)}{r}
\Biggl[ \, 1 + 
\biggl( \frac{\alpha_s(\mu)}{4\pi} \biggr) 
  \biggl\{ 2 \beta_0 \log (\mu' r) + a_1 \biggr\} 
\nonumber \\ && ~~~~~~~~~~~~~~
+
\biggl( \frac{\alpha_s(\mu)}{4\pi} \biggr)^2
  \left\{ \beta_0^2 \Bigl( 4 \log^2 (\mu' r) + \frac{\pi^2}{3} \Bigr)
  + 2 ( \beta_1 + 2 \beta_0 a_1 ) \log (\mu' r) + a_2 \right\}
\Biggr] ,
\label{coulombpot}
\\ &&
\omega = E+i\Gamma_t ,
~~~~~~~~~
E = \sqrt{s}-2 m_t .
\label{omega}
\eea
In above formulas $m_t$ and $\Gamma_t$ denote the pole mass 
and the decay width of top quark, respectively.
$V_{\rm C}(r)$ is the Coulombic part of the $t\bar{t}$ potential 
$V(r)$ including the full second order corrections.
Definitions of all parameters in the above formulas are collected in
Appendix \ref{appa}

Eq.~(\ref{totcs}) includes not only all
${\cal O} (\alpha_s^2, \, \alpha_s \beta, \, \beta^2)$ corrections
to the LO cross section but also,
in the vicinity of each resonance peak, all
${\cal O} (\alpha_s^2)$ corrections to the 
resonance pole position and to its residue.\footnote{
Hereafter we write ${\cal O}(\alpha_s)$, ${\cal O}(\alpha_s^2)$,
etc.\ instead of
${\cal O} (\alpha_s, \, \beta)$,
${\cal O} (\alpha_s^2, \, \alpha_s \beta, \, \beta^2)$, etc.\
for the sake of simplicity.
}
An only difference of eq.~(\ref{totcs})
from the corresponding formula in \cite{mye} is a factor
$i \Gamma_t / 6 m_t$, which arises from a relativistic correction
to the $t\bar{t}$ kinetic energy, 
$\frac{\vc{p}^4}{4m_t^3}$, and from a relativistic correction to
the $t\bar{t}$ production vertex,
$\tilde{\psi}^\dagger \, \sigma^i 
\frac{\stackrel{\leftrightarrow}{\triangle}}{12m_t^2}
\tilde{\chi}$.
This factor is omitted incorrectly in \cite{mye};
numerically its contribution is negligible.\footnote{
The authors of \cite{mye} claim that they incorporate the top quark width
via replacement $E \to E + i\Gamma_t$.
Nevertheless, they do not follow this prescription consistently
in their derivation of $\sigma_{\rm tot}(s)$ and overlook the
factor $i \Gamma_t / 6 m_t$.
}

For $\Gamma_t = 0$, eq.~(\ref{totcs}) becomes independent of
the cutoff $r_0$ as $r_0 \to 0$ up to the order of our interest.
For $\Gamma_t > 0$ there are uncancelled $1/r_0$ and $\log r_0$
singularities due to our improper treatment of $t$ decay processes.
Thus, following \cite{mye} we expand eq.~(\ref{totcs}) in
$r_0$ and omit all terms that vanish as $r_0 \to 0$, and then
we set
\bea
r_0 = 
{e^{2-\gamma_E} \over 2m_t} .
\label{cutoff}
\eea
We also set $m_t=175$~GeV, $\Gamma_t=1.43$~GeV and 
$\alpha_s(m_Z)=0.118$ in our numerical analyses below.
As a cross check of our calculations, we reproduced the total cross sections 
calculated in \cite{yakovlev}.

\begin{figure}[tbp]
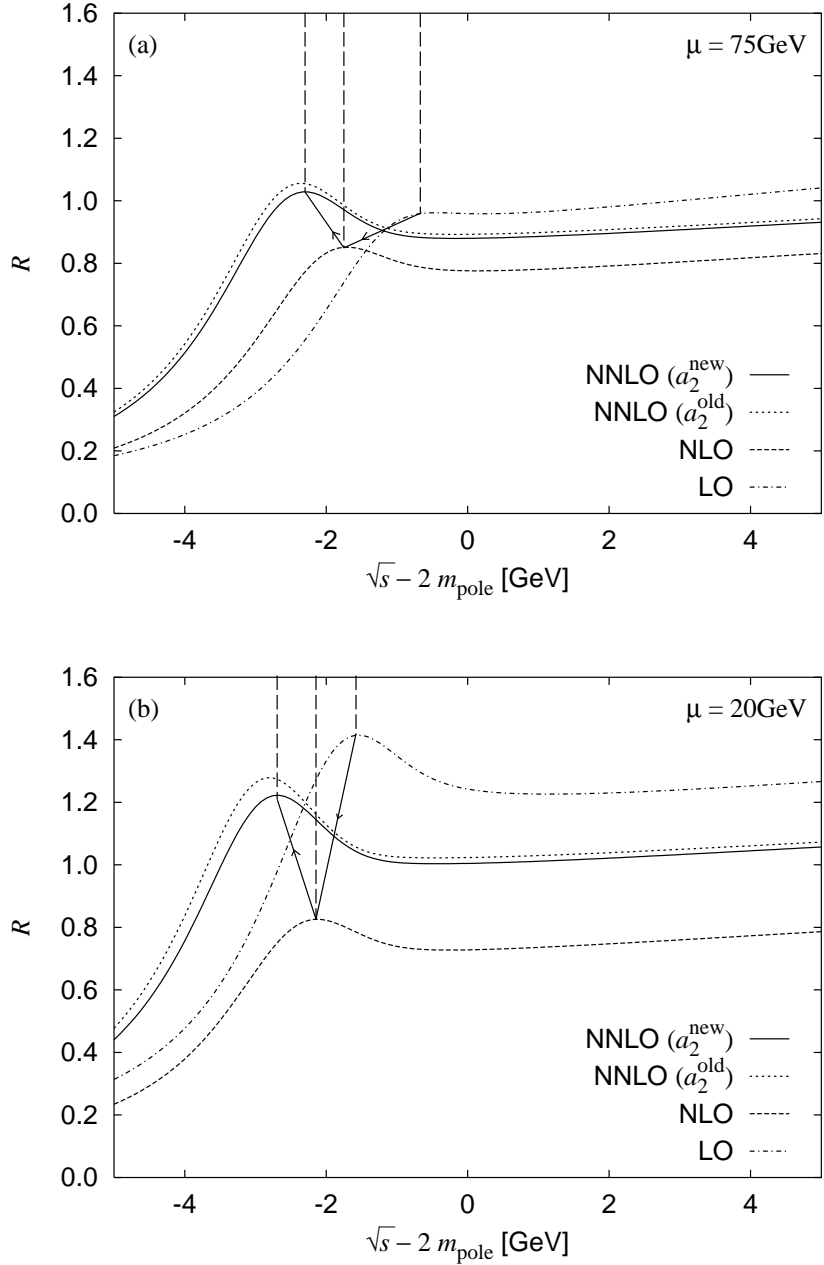

  \begin{center}
    \includegraphics[width=8cm,angle=-90]{R_75_0.eps}\\
    \vspace{8mm}
    \includegraphics[width=8cm,angle=-90]{R_20_0.eps}\\
    \vspace{8mm}
    \caption{\label{fig1}
      $R$-ratios for $e^+ e^- \to t\bar{t}$ 
      at LO (dot-dashed), NLO (dashed), and NNLO (solid) 
      as functions of the energy measured from 
      twice the pole mass, $\sqrt{s}-2 m_{\rm pole}$.
      Arrows indicate dislocations of the maximum point of $R$
      as the ${\cal O}(\alpha_s)$ and ${\cal O}(\alpha_s^2)$
      corrections are included, respectively.
      We set $m_{\rm pole} = m_t = 175$~GeV, 
      $\Gamma_t=1.43$~GeV, and $\alpha_s(m_Z)=0.118$.  
      Dotted lines show NNLO $R$-ratios calculated with an old 
      value of $a_2$ \protect\cite{peter}, which is one of the coefficients 
      in the two-loop 
      perturbative QCD potential.  
      Figure (a) is for $\mu$ = 75~GeV and (b) is for $\mu$ = 20~GeV.  
    }
  \end{center}
\end{figure}

In Figs.~\ref{fig1} we compare the $R$-ratio 
$R(s) = \sigma_{\rm tot}/\sigma_{\rm pt}$ at LO, NLO, and NNLO
($\sigma_{\rm pt} = {4\pi \alpha^2}/{3s}$).
As noted in \cite{ht,mye} the cross 
section changes considerably as we include 
${\cal O}(\alpha_s)$ and ${\cal O}(\alpha_s^2)$ corrections, 
respectively.
One sees that, as we include these corrections, 
convergence of the normalization of the cross section 
is better for $\mu = 75$~GeV than that for $\mu = 20$~GeV, whereas
convergence of the peak position ($\simeq$ mass of the $1S$ resonance)
is better for $\mu = 20$~GeV than that for $\mu = 75$~GeV.
This indicates that the peak position is determined mainly by the
shape of the potential $V(r)$ at the Bohr scale 
$\sim ( \alpha_s m_t)^{-1}$,
while the normalization of the cross section is determined by
the shape of $V(r)$ at a shorter distance;
note that corrections to the
potential are minimized around $r \simeq 1/\mu'=e^{-\gamma_E}/\mu$.
In the same figure we also show the cross section calculated using
an old value \cite{peter} of $a_2$ in $V_{\rm C}(r)$, which has been corrected
recently \cite{schroeder}.  
A change of the cross section caused by correcting $a_2$ is small.

\begin{figure}[tbp]
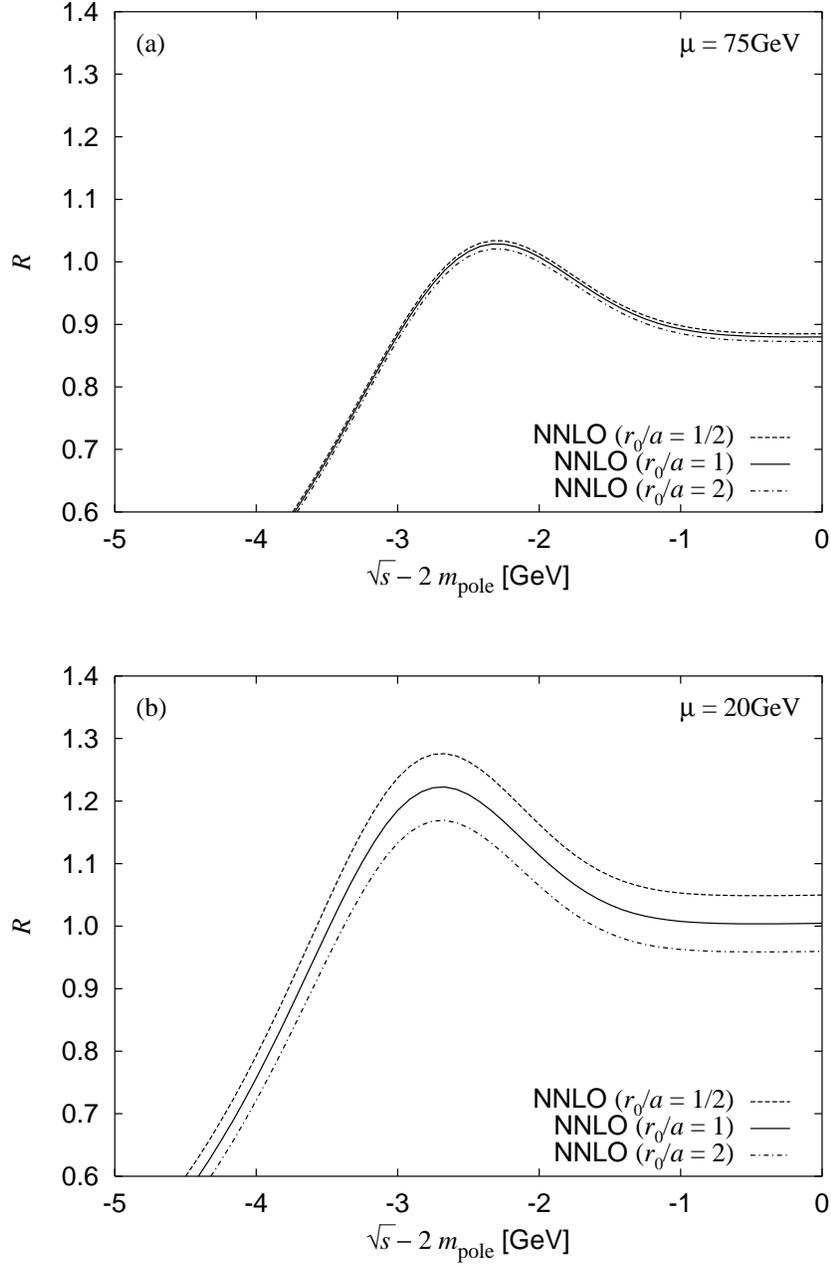

  \begin{center}
    \includegraphics[width=8cm,angle=-90]{R_75_r0.eps}\\
    \vspace{8mm}
    \includegraphics[width=8cm,angle=-90]{R_20_r0.eps}\\
    \vspace{8mm}
    \caption{\label{fig2}
      $R$-ratios for $e^+ e^- \to t\bar{t}$ 
      at NNLO for several values of $r_0$: 
      $r_0$ = $a/2$ (dashed), $r_0=a$ (solid), and $r_0=2a$ (dot-dashed), 
      where $a \equiv {e}^{2-\gamma_{E}}/2m_t$.  
      Figure (a) is for $\mu$ = 75~GeV, and (b) is for $\mu$ = 20~GeV.  
      Other notations and parameters are same as in Fig.~\ref{fig1}.
    }
  \end{center}
\end{figure}

In Figs.~\ref{fig2} we vary the value of $r_0$ by factors 2 and 1/2. 
The cross section varies correspondingly, which is generated by
${\cal O} (\alpha_s^3)$ and ${\cal O} (\frac{\Gamma_t}{m_t})$ terms
in eq.~(\ref{totcs}).
The sizes of the variations serve as a
measure of uncertainties of our theoretical prediction.
They seem to be rather small as compared to what one naively expects from
the poor convergence properties seen in Figs.~\ref{fig1}.

\section{Redefinition of Top Quark Mass}
\label{s3}

According to Beneke \cite{beneke}, 
we define a new quark mass appropriate in the
threshold region
(the potential-subtracted mass) by adding an
infra-red portion of the Coulombic potential to the pole mass.
In this way the new mass is related to 
the $\overline{\rm MS}$ mass in a more convergent
series than to the pole mass (in our case $m_{\rm pole} = m_t$):
\bea
&&
m_{\rm PS}(\mu_f) \equiv m_{\rm pole} + \Delta m (\mu_f) ,
\label{psmass}
\\ &&
\, \Delta m (\mu_f) \equiv \frac{1}{2} 
{\hbox to 18pt{
\hbox to -5pt{$\displaystyle \int$} 
\raise-15pt\hbox{$\scriptstyle |\vc{q}|<\mu_f$} 
}}
{\textstyle \frac{d^3\vc{q}}{(2\pi)^3} }
\, \tilde{V}_{\rm C}(q) ,
\eea
where $\tilde{V}_{\rm C}(q)$ is the Fourier transform of the Coulombic
potential $V_{\rm C}(r)$.\footnote{
Note that our $\Delta m(\mu_f)$ is related to a corresponding quantity
in \cite{beneke} by $\Delta m(\mu_f) = - \delta m (\mu_f)$.
}
At the same time we subtract a corresponding part from the potential
as
\bea
V_{\rm C}(r;\mu_f) \equiv
V_{\rm C}(r) - 2 \Delta m(\mu_f)
\eea
such that the total energy of a quark-antiquark pair remains unchanged in
both schemes:
\bea
2 m_{\rm pole} + V_{\rm C}(r) =
2 m_{\rm PS}(\mu_f) + V_{\rm C}(r;\mu_f) .
\label{totener}
\eea

\begin{figure}[tbp]
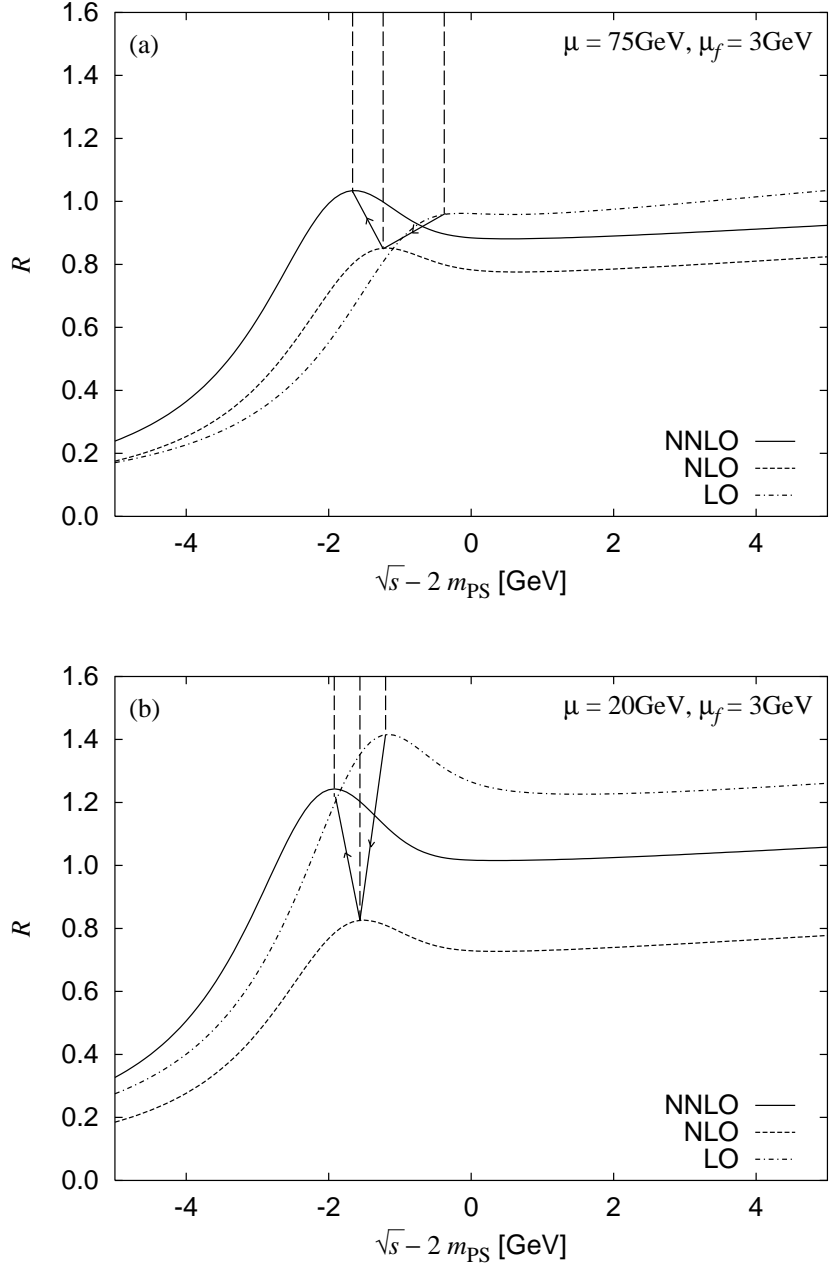

  \begin{center}
    \includegraphics[width=8cm,angle=-90]{R_75_3.eps}\\
    \vspace{8mm}
    \includegraphics[width=8cm,angle=-90]{R_20_3.eps}\\
    \vspace{8mm}
    \caption{\label{fig3}
      $R$-ratios for $e^+ e^- \to t\bar{t}$ 
      at LO (dot-dashed), NLO (dashed), and NNLO (solid) 
      as functions of the energy measured from twice the
      potential-subtracted mass, $\sqrt{s}-2 m_{\rm PS}$.
      We set $\mu_f$ = 3~GeV and $m_{\rm PS}(\mu_f)$ = 175~GeV.  
      Figure (a) is for $\mu$ = 75 GeV, and (b) is for $\mu$ = 20 GeV.  
      Other notations and parameters are same as in Fig.~\ref{fig1}.
    }
  \end{center}
\end{figure}

In Fig.~\ref{fig3} are shown the LO, NLO and NNLO
total cross section by fixing $m_{\rm PS}(3~\mbox{GeV})=175$~GeV.
It can be seen that the convergence of the $1S$ peak position becomes
better as expected.
Meanwhile the normalization of the cross section scarcely changes by
this modification.
It is because eq.~(\ref{psmass}) essentially incorporates a 
constant shift of 
the cross section in the horizontal direction by an amount $\Delta m(\mu_f)$, 
while changes in the normalization 
generated by a modification of the mass in 
the Schr\"odinger equation (\ref{green}) is negligibly small.

\section{Renormalization-Group Improvement of $V_{\rm C}(r)$}
\label{s4}

As already mentioned, it is important to resum logarithms in
calculations of threshold cross sections.
We demonstrate\footnote{
A full resummation of logarithms up to NNLO requires a significant
modification of the formulas (\ref{totcs}) and (\ref{green});
we will study its incorporation in our future work.
} 
an improvement of convergence of the cross section
by incorporating log resummations to the Coulombic
potential $V_{\rm C}(r)$.

The Coulombic potential $V_{\rm C}(r)$
is identified with the QCD potential between a static
quark-antiquark pair.
If we write this potential in momentum space 
(Fourier transform of eq.~(\ref{coulombpot})) as
\bea
\tilde{V}_{\rm C} (q) = - 4\pi C_F \frac{\alpha_V(q;\mu)}{q^2} ,
\eea
a log resummation using a renormalization group equation
is achieved simply by a replacement $\mu \to q$ \cite{peter}:
\bea
\tilde{V}_{\rm C}^{(RG)} (q) = - 4\pi C_F \frac{\alpha_V(q;q)}{q^2} .
\label{RGimp}
\eea
Hence, in accordance with the formulation in the previous section,
we define a potential-subtracted mass and
a renormalization-group-improved potential in coordinate
space, respectively, as
\bea
&&
m_{\rm PS}(\mu_f) \equiv m_{\rm pole} + \Delta m (\mu_f) ,
~~~~~~~~
\Delta m (\mu_f) \equiv \frac{1}{2} 
{\hbox to 18pt{
\hbox to -5pt{$\displaystyle \int$} 
\raise-15pt\hbox{$\scriptstyle |\vc{q}|<\mu_f$} 
}}
{\textstyle \frac{d^3\vc{q}}{(2\pi)^3} }
\, \tilde{V}_{\rm C}^{(RG)}(q) ,
\\ &&
V_{\rm C}^{(RG)}(r;\mu_f) \equiv
{\hbox to 18pt{
\hbox to -5pt{$\displaystyle \int$} 
\raise-15pt\hbox{$\scriptstyle |\vc{q}|>\mu_f$} 
}}
{\textstyle \frac{d^3\vc{q}}{(2\pi)^3} }
\, e^{i \vc{q} \cdot \vc{r}} \, \tilde{V}_{\rm C}^{(RG)}(q) 
\nonumber \\ && ~~~~~~~~~~~~~~~~
= V_{\rm C}^{(RG)}(r;\mu_f=0) -
{\hbox to 18pt{
\hbox to -5pt{$\displaystyle \int$} 
\raise-15pt\hbox{$\scriptstyle |\vc{q}|<\mu_f$} 
}}
{\textstyle \frac{d^3\vc{q}}{(2\pi)^3} }
\, e^{i \vc{q} \cdot \vc{r}} \, \tilde{V}_{\rm C}^{(RG)}(q) 
 .
\label{rdepsubt}
\eea
In this formulation both $m_{\rm pole}$ and 
$\Delta m (\mu_f)$ suffer from theoretical uncertainties of the order
$\sim \Lambda_{\rm QCD}$ due to the renormalon poles, but they
cancel in $m_{\rm PS}(\mu_f)$.
We note that strictly speaking
there is no guiding principle for subtracting also a
$r$-dependent part from the potential in (\ref{rdepsubt}), since there
is no known renormalon cancellation related to $r$-dependent part of 
the potential.
In fact the total energy of a quark-antiquark pair
(\ref{totener}) is
not well-defined after the renormalization-group improvement 
(\ref{RGimp}), and a theoretical ambiguity of the order
$\sim \Lambda_{\rm QCD}^2 r$ is caused by a non-cancelled renormalon pole
in the $r$-dependent part.\footnote{
Within our perturbative
formalism $\sim \Lambda_{\rm QCD}^2 r$
term in the potential is forbidden by the rotational invariance,
and the first ambiguous $r$-dependence arises
at $\sim \Lambda_{\rm QCD}^3 r^2$.
}
This ambiguity is negligible in our case thanks to the large mass and
decay width of the top quark \cite{fk}; see \cite{sp,fy,sfhmn} for
more practical analyses.
Thus, we should set $\mu_f \gg \Lambda_{\rm QCD}$ in order to avoid a
bad convergence of the cross section 
generated by a renormalon pole, while we should
set $\mu_f \ll \alpha_s m_t$ such that a main part of bound-state
dynamics is preserved.
In our analyses below we choose
$\mu_f = 3$~GeV.
(We have checked that upon varying $\mu_f$ the cross section changes 
only by a constant shift in the horizontal direction and a change in
the normalization is negligible,
i.e.\ $r$-dependence of the subtracted part in (\ref{rdepsubt}) 
plays no significant role.)
\footnote{
In rewriting the pole mass $m_t$ in terms of $m_{\rm PS}(\mu_f)$
in eqs.~(\ref{totcs})-(\ref{omega}),
we retained terms up to (and including)
${\cal O}(\alpha_s^3)$ in this relation.
}

\begin{figure}[tbp]
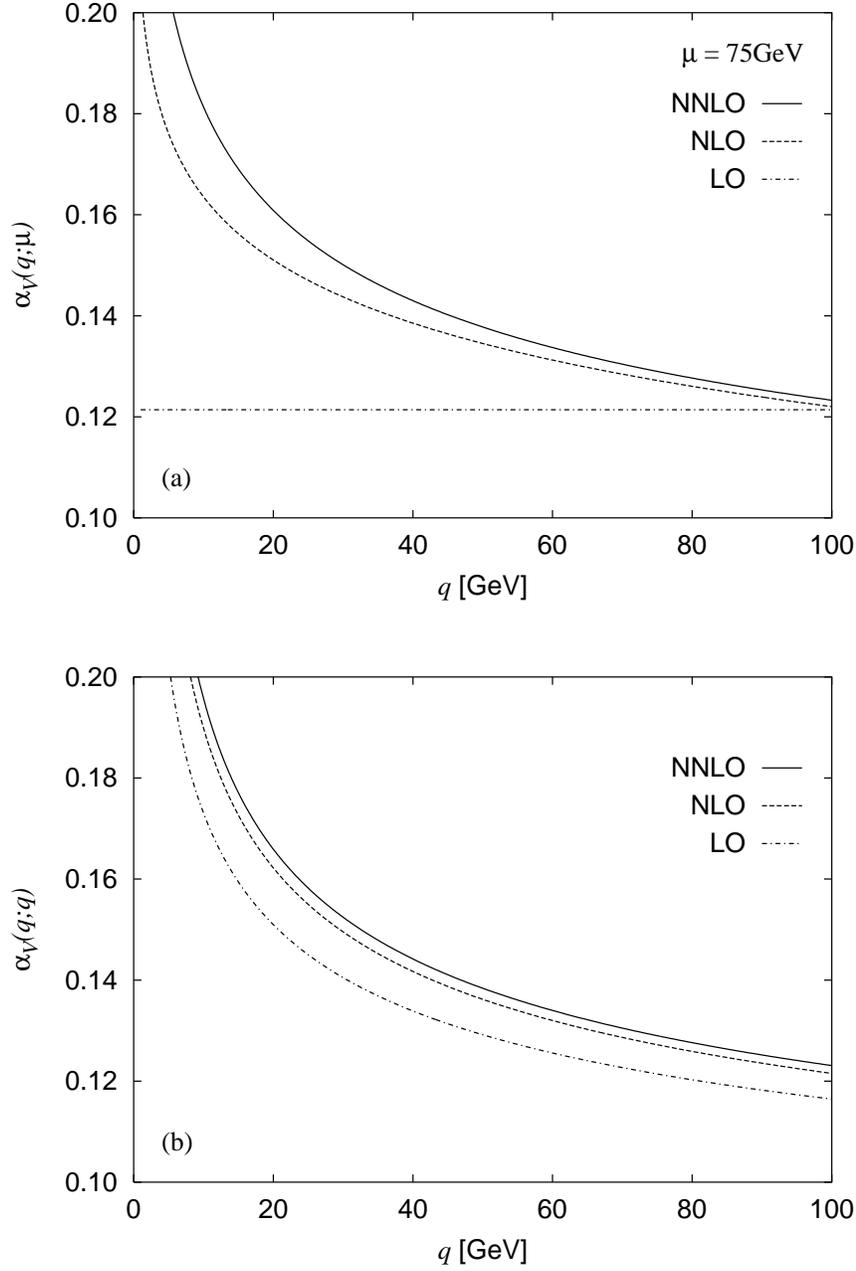

  \begin{center}
    \includegraphics[width=8cm,angle=-90]{aV_75.eps}\\
    \vspace{8mm}
    \includegraphics[width=8cm,angle=-90]{aV_q.eps}\\
    \vspace{8mm}
    \caption{\label{fig4}
      The momentum-space couplings $\alpha_V$ vs.\ momentum transfer
      $q$ at LO (dot-dashed), NLO (dashed), and 
      NNLO (solid).
      Figure (a) is the fixed-order coupling ($\mu$ = 75 GeV), and (b)
      is a renormalization-group improved coupling ($\mu$ = $q$).  
    }
  \end{center}
\end{figure}

\begin{figure}[tbp]
  \begin{center}
    \includegraphics[width=8cm,angle=-90]{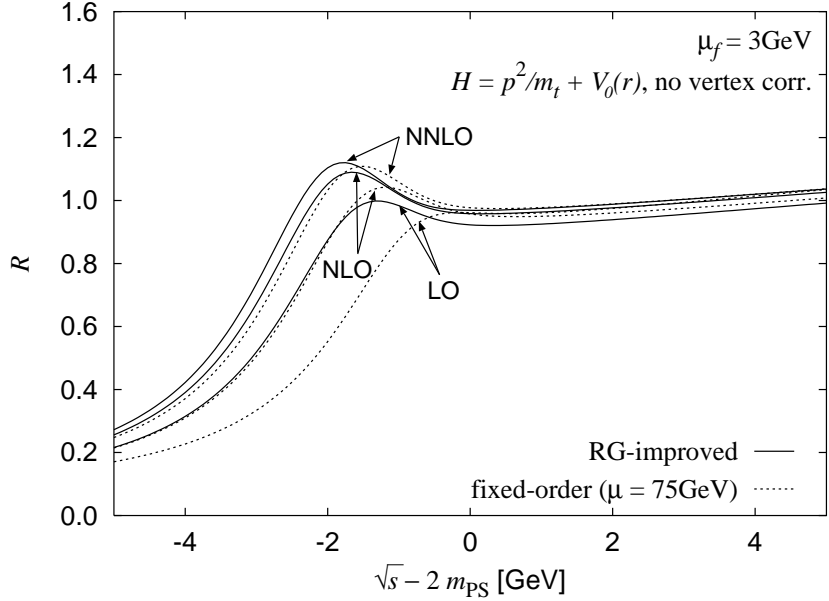}\\
    \vspace{8mm}
    \caption{\label{fig5}
      $R$-ratios for $e^+ e^- \to t\bar{t}$ 
      calculated with a Hamiltonian 
      $H = p^2/m_t + V_0(r)$, where $V_0(r)$ includes 
      only the Coulombic part of the $t\bar{t}$ potential.  
      Other corrections (vertex renormalization constants, 
      kinematical corrections, etc.)\ are not included.
      Solid and dashed lines, respectively, show $R$-ratios with 
      ($V_0(r)=V_{\rm C}^{(RG)}(r;\mu_f)$) and
      without ($V_0(r)=V_{\rm C}(r)$, $\mu=75$~GeV) a
      renormalization-group improvement of the Coulombic potential.
      We set $\mu_f$ = 3~GeV, $m_{\rm PS}(\mu_f)$ = 175~GeV,
      $\Gamma_t=1.43$~GeV, and $\alpha_s(m_Z)=0.118$.    
    }
  \end{center}
\end{figure}

We compare the couplings of the momentum-space potential with
[$\alpha_V(q;q)$] and without 
[$\alpha_V(q;\mu=75~\mbox{GeV})$] a renormalization-group improvement
in Figs.~\ref{fig4}.
One sees that convergence of the coupling improves drastically 
by the log resummation over the whole range of our interest,
$m_t^{-1} < r \simlt (\alpha_s m_t)^{-1}$.
One therefore anticipate that ${\cal O}(\alpha_s)$ and 
${\cal O}(\alpha_s^2)$ corrections to the total cross section
originating from $V_{\rm C}(r)$
also become smaller and more converging.
In order to see only these corrections separately, we show in Fig.~\ref{fig5}
the $R$-ratio calculated from 
\bea
R(s) = \frac{6\pi N_c Q_t^2}{m_t^2} \, \mbox{Im} \, G(0,0) 
\eea
with
\bea
\left\{
-\frac{1}{m_t} 
\left[ \frac{d^2}{dr^2}+\frac{2}{r}\frac{d}{dr} \right]
+ V_0(r) - \omega
\right\} G(r,r') = \frac{1}{4\pi r r'} \, \delta (r-r') ,
\eea
both for $V_0(r)=V_{\rm C}(r)$ and $V_0(r)=V_{\rm C}^{(RG)}(r;\mu_f)$.
Namely, we omit all ${\cal O}(\alpha_s)$ and
${\cal O}(\alpha_s^2)$ corrections other than those in the Coulombic
potential.
One sees clearly that the convergence property has improved 
considerably by the log resummations.

\begin{figure}[tbp]
  \begin{center}
    \includegraphics[width=8cm,angle=-90]{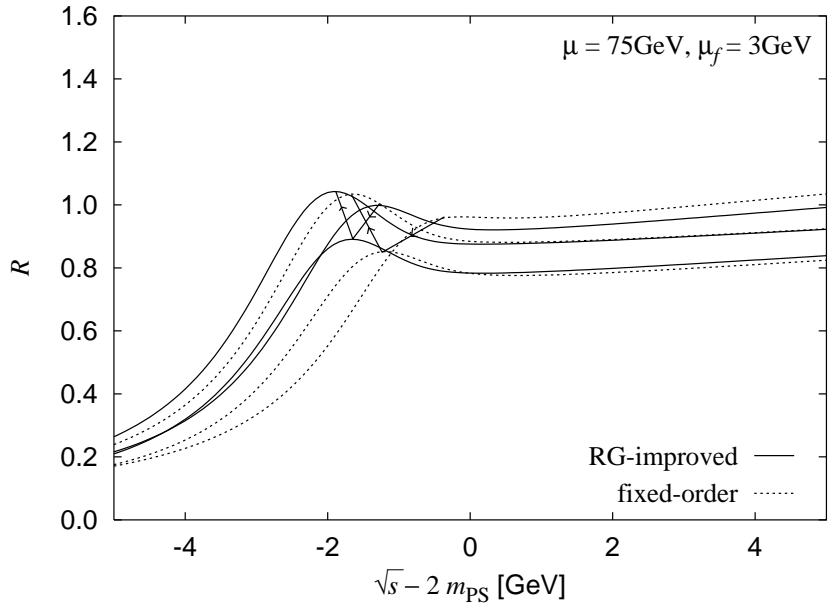}\\
    \vspace{8mm}
    \caption{\label{fig6}
      $R$-ratios for $e^+ e^- \to t\bar{t}$ 
      at LO, NLO, and NNLO.
      Solid lines show those with renormalization-group improved
      Coulombic potentials, $V_{\rm C}^{(RG)}(r;\mu_f)$.
      Dashed lines are those with fixed-order Coulombic potentials
      $V_{\rm C}(r)$.
      Arrows indicate dislocations of the maximum point of $R$
      as the ${\cal O}(\alpha_s)$ and ${\cal O}(\alpha_s^2)$
      corrections are included, respectively.
      We set $\mu_f$ = 3~GeV, $m_{\rm PS}(\mu_f)$ = 175~GeV, $\mu=75$~GeV,
      $\Gamma_t=1.43$~GeV, and $\alpha_s(m_Z)=0.118$.    
    }
  \end{center}
\end{figure}

Finally we combine the above corrections with all other corrections.
Namely we show in Fig.~\ref{fig6} the total cross section 
(\ref{totcs}) with and without the renormalization-group improvement
of the Coulombic potential.
Also we list the ``binding energies'' of the $1S$ resonance state
$2m_{\rm PS}(\mu_f)-M_{\rm 1S}$ in Table~\ref{table}.
\begin{table}[tbp]
\begin{center}
\begin{minipage}{11cm}
\begin{tabular}{l|cccc} \hline\hline
&
\multicolumn{2}{c}{(fixed-order)} &
\multicolumn{2}{c}{(RG-improved)}
\\
      & $\mu=20$~GeV     & $\mu=75$~GeV
      & $\mu=20$~GeV     & $\mu=75$~GeV
\\ \hline
LO    & $1.390$~GeV                     & $0.838$~GeV
      & $1.573$~GeV                     & $1.573$~GeV  
\\
NLO   & $1.716$~GeV                     & $1.453$~GeV
      & $1.861$~GeV                     & $1.861$~GeV
\\
NNLO  & $2.062$~GeV                     & $1.817$~GeV
      & $2.136$~GeV                     & $2.058$~GeV
\\ \hline\hline
\end{tabular}
\end{minipage}
\end{center}
\caption{\label{table}
``Binding energies'' of the $1S$ resonance state
defined as $2m_{\rm PS}(\mu_f)-M_{1S}$ at LO, NLO, and 
NNLO calculated with $V_{\rm C}(r)$ (fixed-order) and with 
$V_{\rm C}^{(RG)}(r;\mu_f)$ (RG-improved).
We set $\mu_f$ = 3~GeV, $m_{\rm PS}(\mu_f)$ = 175~GeV, 
$\Gamma_t=1.43$~GeV, and $\alpha_s(m_Z)=0.118$.    
}
\end{table}
Although it is seen that convergence of the normalization of the cross 
section as well as convergence of the
$1S$ resonance mass become slightly better, improvements 
are not so drammatic.
This is because other corrections, in particular those originating from 
the $1/r^2$ potential in $V(r)$, are uncomfortably large.
It remains as our future task to gain better understandings of 
these residual large corrections.

\section{Top Quark Momentum Distribution}
\label{s5}

Using the NRQCD formalism and also techniques developed in \cite{mye},
one obtains the momentum distribution of top quarks in the threshold
region including all ${\cal O}(\alpha_s^2)$ corrections as
\bea
&&
\frac{d\sigma}{dp} = \frac{16\alpha^2}{s^2} \,
N_c Q_q^2 \, 
\left\{
1 
+ \left( \frac{\alpha_s(m_t)}{\pi} \right) C_F C_1
+ \left( \frac{\alpha_s(m_t)}{\pi} \right)^2 C_F C_2(r_0)
\right\}
\times p^2 \Gamma_t  \,  f(p;r_0) ,
\label{momdist}
\eea
where
\bea
&&
f(p;r_0) = \left\{
\left( 1 + \frac{2E}{3m_t} \right) | \tilde{G}(p;r_0) |^2 
+ \frac{3}{2} C_F \alpha_s(\mu)^2 \,
\mbox{Re} \left[
\tilde{G}_{1/r}(p;r_0) \, \tilde{G}(p;r_0)^*
\right]
\right.
\nonumber
\\ && ~~~~~~~~~~ \left.
-\frac{11}{6}C_F \alpha_s(\mu)^2  \,
\mbox{Re} \left[
\tilde{G}_{ip_r}(p;r_0) \, \tilde{G}(p;r_0)^*
\right]
+ \frac{1}{6m_t} \, \frac{\sin (pr_0)}{pr_0} \, 
\mbox{Re} \left[
\tilde{G}(p;r_0) 
\right] \right\} .
\eea
In these formulas, $p$ denotes the magnitude of the top quark three-momentum.
Momentum-space Green functions are defined from the coordinate-space
Green function
in (\ref{green}) by
\bea
&&
\tilde{G}(p;r_0) =
\int d^3 \vc{r} \, e^{i \vc{p} \cdot \vc{r}} \, G(r,r_0) ,
\label{gp1}
\\ &&
\tilde{G}_{1/r}(p;r_0) =
\int d^3 \vc{r} \, e^{i \vc{p} \cdot \vc{r}} \, 
\frac{1}{\alpha_s(\mu) m_t r}\,  G(r,r_0) ,
\label{gp2}
\\ &&
\tilde{G}_{ip_r}(p;r_0) =
\int d^3 \vc{r} \, e^{i \vc{p} \cdot \vc{r}} \, 
\frac{i p_r}{\alpha_s(\mu) m_t}  \,  G(r,r_0) ,
\label{gp3}
\eea
with $ip_r = d/dr + 1/r$.
A derivation of the formulas is given in Appendix \ref{appextra}.
One can show that upon integrating over $\int dp$ the total cross section
formula (\ref{totcs}) is recovered.
A proof of the unitarity relation between the total cross section
(\ref{totcs}) and the momentum distribution (\ref{momdist}) is given 
in Appendix \ref{appb}.
We also checked numerically that the unitarity relation holds
well within our desired accuracies.

For consistency with our analyses of the total cross section, we
expand eq.~(\ref{momdist}) in terms of the cutoff 
$r_0$, omit terms regular as $r_0 \to 0$, and set its value 
as in eq.~(\ref{cutoff}).\footnote{
Note that strictly speaking the unitarity relation is violated
after this expansion,
because $\int dp$ integration and expansion in $r_0$ do not
commute for $\Gamma_t > 0$.
Practically the unitarity relation holds to a sufficient accuracy by
cuting off the momentum integration at some appropriately
large scale.
}
In all figures we choose $\mu=20$~GeV since a relevant scale around
the distribution peak is the scale of Bohr radius 
$\sim (\alpha_s m_t)^{-1}$.

\begin{figure}[tbp]
  \begin{center}
    \includegraphics[width=8cm,angle=-90]{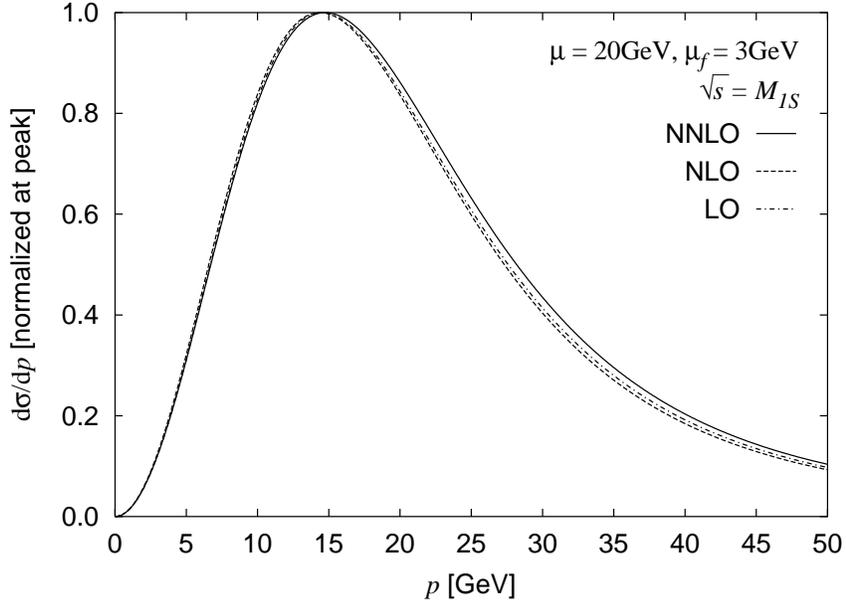}\\
    \vspace{8mm}
    \caption{\label{fig7}
      Top quark momentum distributions at LO (dot-dashed), NLO (dashed), 
      and NNLO (solid) for $\mu$ = 20~GeV.  
      For each curve, we set the c.m.\ energy on the $1S$ resonance
      state, $\sqrt{s}=M_{1S}$.
    }
  \end{center}
\end{figure}

\begin{figure}[tbp]
  \begin{center}
    \includegraphics[width=8cm,angle=-90]{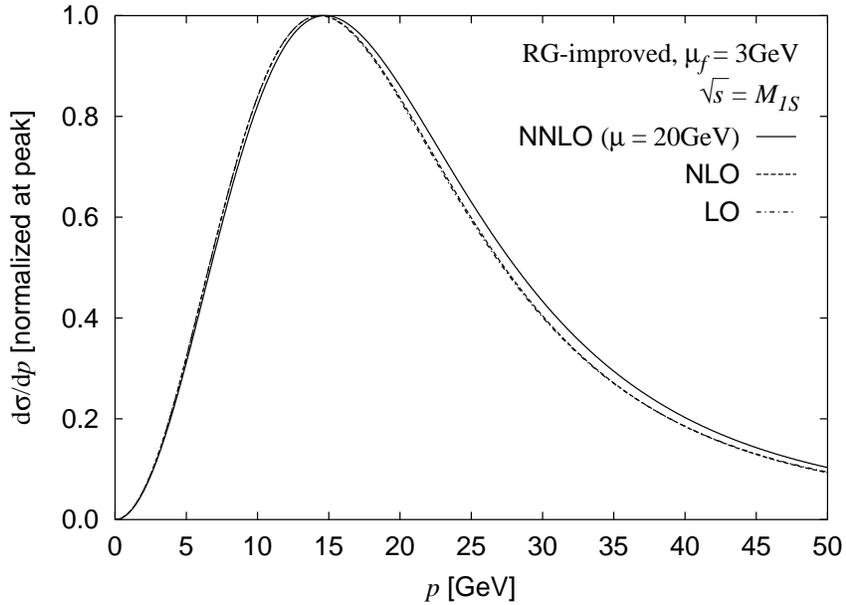}\\
    \vspace{8mm}
    \caption{\label{fig8}
      Same as Fig.~\protect\ref{fig7} but with a renormalization
      group improvement in the Coulomb part of the potentail:
      LO (dot-dashed), NLO (dashed), and NNLO (solid). 
    }
  \end{center}
\end{figure}

\begin{figure}[tbp]
  \begin{center}
    \includegraphics[width=8cm,angle=-90]{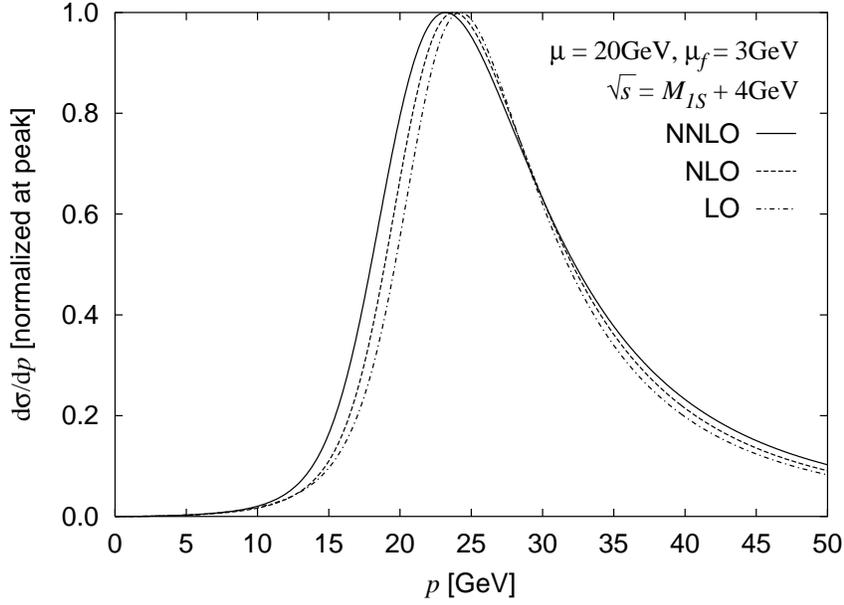}\\
    \vspace{8mm}
    \caption{\label{fig9}
      Top quark momentum distributions at LO (dot-dashed), NLO (dashed), 
      and NNLO (solid) for $\mu$ = 20~GeV.  
      For each curve, we set the c.m.\ energy at 4~GeV above the 
      $1S$ resonance mass.
    }
  \end{center}
\end{figure}

\begin{figure}[tbp]
  \begin{center}
    \includegraphics[width=8cm,angle=-90]{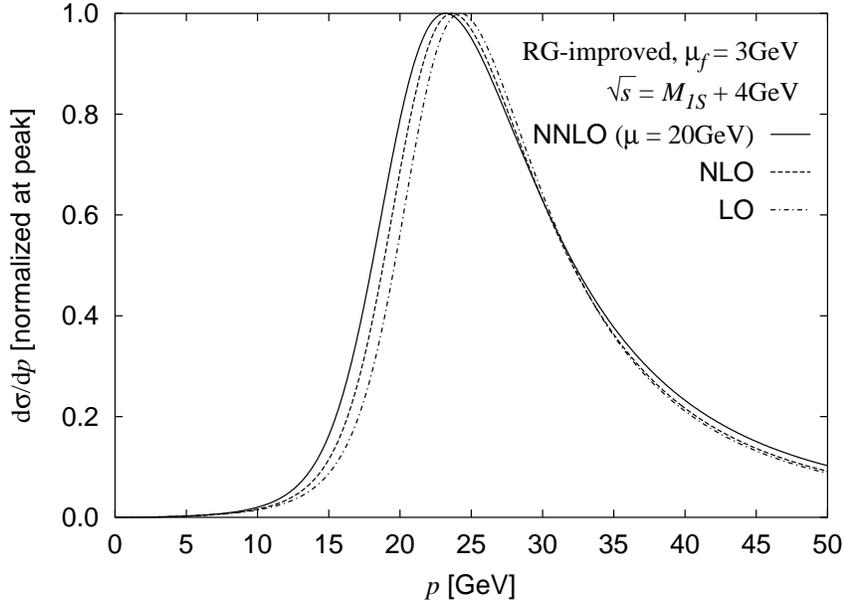}\\
    \vspace{8mm}
    \caption{\label{fig10}
      Same as Fig.~\protect\ref{fig9} but with a renormalization
      group improvement in the Coulomb part of the potentail:
      LO (dot-dashed), NLO (dashed), and NNLO (solid). 
    }
  \end{center}
\end{figure}

Top quark momentum distributions 
(normalized to unity at each distribution peak)
are shown in Figs.~\ref{fig7}-\ref{fig10}.
Following a strategy advocated in \cite{fms},
we fix the c.m.\ energy relative to the $1S$ resonance mass 
$\Delta E = \sqrt{s} - M_{1S}$ upon comparing 
LO, NLO and NNLO distributions.
On the $1S$ resonance ($\Delta E =0$, Fig.~\ref{fig7}), 
${\cal O}(\alpha_s)$ and
${\cal O}(\alpha_s^2)$ corrections shift the distribution peak,
$p_{\rm peak}$,
by $-0.8\%$ and by $+2.5\%$, respectively.
Also one sees that the ${\cal O}(\alpha_s^2)$ corrections are larger
at higher momentum region.
This is as expected because part of the ${\cal O}(\alpha_s^2)$ 
corrections are relativistic corrections which
are enhanced in the relativistic regime.
In Fig.~\ref{fig8} we incorporate a log resummation in the Coulombic
potential, i.e.\ replace $V_{\rm C}(r) \to V_{\rm C}^{(RG)}(r;\mu_f)$.
Qualitative tendencies of the corrections are not changed by the
resummation.
($\delta p_{\rm peak}/p_{\rm peak} = +0.5\%$ and $+2.2\%$ at 
${\cal O}(\alpha_s)$ and ${\cal O}(\alpha_s^2)$, respectively.)
We show momentum distributions at 
$\Delta E = 4$~GeV in Fig.~\ref{fig9} (with $V_{\rm C}(r)$)
and in Fig~\ref{fig10} (with $V_{\rm C}^{(RG)}(r;\mu_f)$).
One sees that in both figures
${\cal O}(\alpha_s)$ and ${\cal O}(\alpha_s^2)$ corrections,
respectively, reduce the peak momentum $p_{\rm peak}$.

In general, we see following
energy dependences of the ${\cal O}(\alpha_s)$ and 
${\cal O}(\alpha_s^2)$ corrections 
to the peak momentum
$\delta p_{\rm peak}/p_{\rm peak}$.
At $\Delta E =0$ the corrections are positive $\sim + \, \mbox{few} \, \%$;
between $\Delta E =0$ and $\Delta E = 1$-2~GeV, the corrections
decrease and change sign from $+ \, \mbox{few} \, \%$ to 
$- \, \mbox{few} \, \%$;
at higher energies, $\Delta E > 1$-2~GeV,
the corrections stay negative, but their 
magnitude $|\delta p_{\rm peak}/p_{\rm peak}|$  
decrease with energy.
The energy dependences of the 
${\cal O}(\alpha_s)$ and 
the ${\cal O}(\alpha_s^2)$ 
corrections are
qualitatively similar.

These energy dependences can be understood
as a consequence of an increase of 
attractive force between $t$ and $\bar{t}$.\footnote{
In fact the strength of the Coulombic force,
$|dV_{\rm C}/dr|$ or $|dV_{\rm C}^{(RG)}/dr|$, increases by the
${\cal O}(\alpha_s)$ and ${\cal O}(\alpha_s^2)$ corrections
at relevant distances.
(This may be seen from increases of the couplings in Fig.~\ref{fig4}.)
Also, there is an additional attractive force 
($1/r^2$ term in $V(r)$) at NNLO.
Thus, reflecting the increase of binding energies,
the mass of the $1S$ resonance state decreases; see Table~\ref{table}.
}
Namely, at $\Delta E =0$, $p_{\rm peak}$ is determined
by the binding energy and is larger for a larger binding energy.
At higher energies, $\Delta E > 1$-2~GeV, 
the peak momentum of the distribution
tends to be determined only from kinematics,
$p_{\rm peak} \approx \frac{1}{2} \sqrt{s-4m_t^2}$.
Meanwhile, if the binding energy becomes larger due to an increase
of attractive force, the $1S$ resonance mass will be lowered,
and therefore $\sqrt{s}$ becomes smaller for a fixed $\Delta E$.

\begin{figure}[tbp]
  \begin{center}
    \includegraphics[width=8cm,angle=-90]{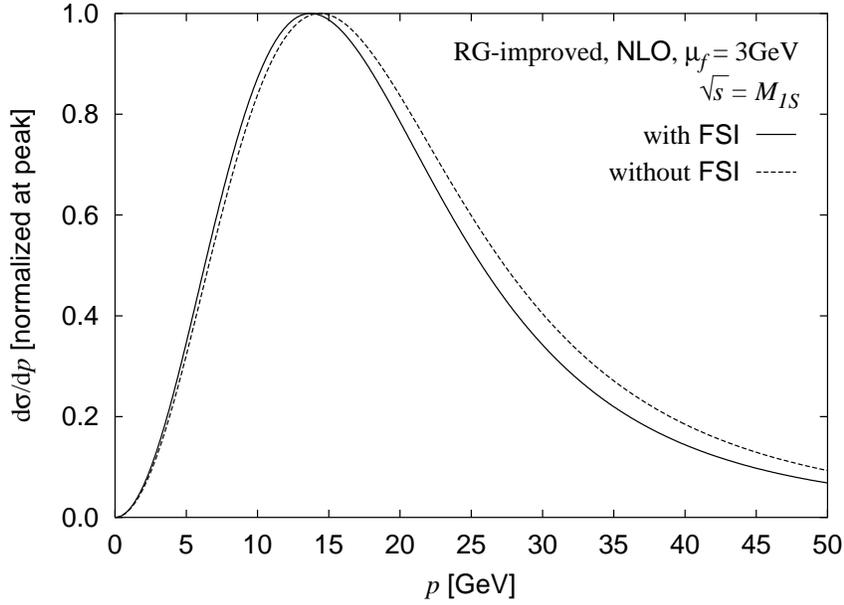}\\
    \vspace{8mm}
    \caption{\label{fig11}
      Top quark momentum distributions at NLO with the renormalization
      group improvement for the Coulomb part of the potential.
      The c.m.\ energy is set on the $1S$ resonance state.
      The solid (dashed) line is calculated with (without) the
      ${\cal O}(\alpha_s)$ final-state interaction corrections.
    }
  \end{center}
\end{figure}

\begin{figure}[tbp]
  \begin{center}
    \includegraphics[width=8cm,angle=-90]{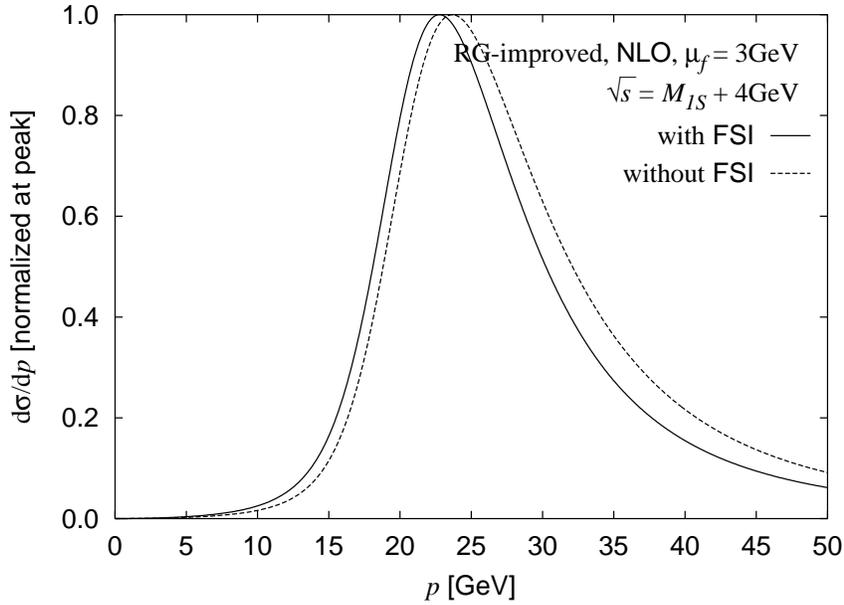}\\
    \vspace{8mm}
    \caption{\label{fig12}
      Same as Fig.~\protect\ref{fig11} but for the c.m.\
      energy 4~GeV above the $1S$ resonance state.
    }
  \end{center}
\end{figure}

In all above results, the decay process of top quarks have been treated 
only effectively by the replacement (\ref{repl}), and we have not 
included in our analyses even the already
known ${\cal O}(\alpha_s)$ corrections which arise in relation to the 
top decay process, namely the final-state
interactions between $t$ and $\bar{t}$ decay products.
For comparison, we show in Figs.~\ref{fig11} and \ref{fig12} 
these effects of the ${\cal O}(\alpha_s)$ final-state interactions 
on the top quark momentum distribution.
As noted in \cite{thesis,cracow,fms,hjkp,ps}, 
the final-state interactions reduce
the peak momentum about 5\% almost independently of the energy.
These energy dependences are distinctly different from those of 
the NLO and NNLO corrections studied above.
Thus, the effects of the ${\cal O}(\alpha_s)$
final-state interactions are larger and
qualitatively different, so that they would be distinguishable from
other NLO and NNLO corrections considered in this paper.

\section{Summary and Discussion}
\label{s6}

We studied convergence properties of the
total cross section for 
$e^+e^- \to t\bar{t}$ in the threshold region.
By expressing the cross section in terms of the
potential-subtracted mass $m_{\rm PS}(\mu_f)$ instead of the pole mass,
a better convergence of the $1S$ resonance mass was obtained,
whereas the normalization of the cross section hardly changed.
We argue that log resummations are indispensable for analyses of
the cross section in the threshold region.
As a first step, we resummed logarithms in the Coulombic part
of the $t\bar{t}$ potential by renormalization-group improvement. 
In this prescription, we followed closely a formulation of the
potential subtraction in the fixed-order analysis.
Corrections originating from the Coulombic potential
became much more converging after the log resummations,
both for
the $1S$ resonance mass and for the normalization of the cross section.
There still remain, however, unexpectedly large 
${\cal O}(\alpha_s^2)$ corrections, whose main part arises
from the $1/r^2$ term in the $t\bar{t}$ potential $V(r)$.
We should implement full log resummations to the threshold cross section
and see whether these large corrections remain.

We also calculated the momentum distributions of top quarks in
the threshold region including full 
${\cal O}(\alpha_s^2)$ corrections.
On the $1S$ resonance state, the ${\cal O}(\alpha_s^2)$ corrections
to the distribution shape are small.
In particular the shift of $p_{\rm peak}$ is $+2.2\%$ after
a renormalization-group improvement of the Coulombic potential,
which seems to be of a legitimate size.
At higher energies, the corrections change sign and become negative.
Over the whole threshold region the size of the corrections 
$\delta p_{\rm peak}/p_{\rm peak}$ stays
within a few \%.
These features can be understood as a combined effect of
kinematics and an increase of binding energy.
Thus, again major part of the corrections can be traced back to
the $1/r^2$ term in $V(r)$ which
affects the binding energy significantly.
Besides full resummations of logarithms,
it is mandatory to incorporate the decay process of top quarks
properly in order to attain a more reliable 
theoretical prediction of the momentum 
distributions, since off-shell contributions, i.e.\
$\sim (p - p_{\mbox{\scriptsize on-shell}})^2/m_t^2$ corrections,
are not treated correctly in the present calculation.
We demonstrated that the ${\cal O}(\alpha_s)$ 
final-state interaction corrections to the distribution shape
are significant in comparison to other NLO corrections.
Thus, we think that yet uncalculated ${\cal O}(\alpha_s^2)$
final-state interactions may give rise to corrections 
which are non-negligible
compared to the NNLO corrections calculated in this paper.

It was argued in \cite{jkpst} that a large theoretical uncertainty
exists even
after a renormalization-group improvement of the Coulombic potential.
This claim was based on a large discrepancy between results of
renormalization-group improvements in momentum space and in coordinate space.
Now we have a better guiding principle.
The large discrepancy originated from a renormalon pole
\cite{jps,beneke}, and by
adopting an appropriate mass definition we can
cancel this pole (at least
in the $r$-independent part of the Coulombic potential) and obtain
a more convergent perturbative series consequently.
In this work, we adopted the potential-subtracted mass.

After completion of this work, we received a paper
by Beneke, Signer and Smirnov \cite{bss}.
Their work has a significant overlap with Section~\ref{s3} of 
the present paper.
Effects of introducing $m_{\rm PS}(\mu_f)$ on the cross section
are consistent between their results and ours.
We adopt a value of $\mu_f$ considerably smaller than that adopted 
in their paper.
This is in view of our
application of the formalism to the renormalization-group improved potential;
see discussion below eq.~(\ref{rdepsubt}).

\section*{\bf Acknowledgements}

The authors are grateful to T.~Teubner for bringing our 
attention to \cite{schroeder}.
One of the authors (Y.S.)\ would like to thank K.~Fujii, K.~Melnikov,
A.~Hoang and T.~Teubner
for useful discussions.
This work is supported by the Japanese-German Cooperative Science
Promotion Program.

\begin{appendix}

\section{Definitions and Conventions}
\label{appa}

In eq.~(\ref{totcs}), the 
vertex renormalization constants are given by \cite{ht,mye}
\bea
C_1=-4,
~~~~~~~~~~~~~
C_2 = C_F \, C_2^A + C_A \, C_2^{NA} + T_R N_L \, C_2^L
      + T_R N_H \, C_2^H ,
\eea
where
\bea
&&
C_2^A = 
\frac{39}{4}-\zeta_3
+\pi^2 \left\{ \frac{2}{3} \log (2 e^{\gamma_E-2} m_t r_0)
+ \frac{4}{3}\log 2 - \frac{35}{18} \right\} ,
\\ &&
C_2^{NA} =
-\, \frac{151}{36}-\frac{13}{2}\zeta_3
+\pi^2 \left\{ \log (2 e^{\gamma_E-2} m_t r_0)
- \,\frac{8}{3}\log 2 + \frac{179}{72} \right\} ,
\\ &&
C_2^L = 
\frac{11}{9} ,
\\ &&
C_2^H =
\frac{44}{9}-\frac{4}{9}\pi^2 .
\eea
QCD color factors are defined as 
$N_c=3$, $C_F=4/3$, $C_A=3$, $T_R=1/2$, and the fermion numbers in
our problem are given by $N_L=5$ and $N_H=1$.
Also, the top quark charge is defined by $Q_t=2/3$.

The Coulombic potential (\ref{coulombpot}) is identified with the
QCD potential between a static quark-antiquark pair.
The first-order correction to the QCD potential was calculated in
\cite{fischler,billoire}, while the second-order correction was 
calculated first in \cite{peter}, a part of which has been corrected 
recently in \cite{schroeder}.
Their coefficients are given, respectively, by
\bea
&&
\beta_0 = \frac{11}{3}C_A - \frac{4}{3} T_R N_L ,
\\ &&
\beta_1 = \frac{34}{3}C_A^2 - \frac{20}{3}C_A T_R N_L 
- 4 C_F T_R N_L ,
\\ &&
a_1 = \frac{31}{9}C_A - \frac{20}{9}T_R N_L ,
\\ &&
a_2 = 
\left( \frac{4343}{162} + 4\pi^2-\frac{\pi^4}{4}+\frac{22}{3}\zeta_3
\right) C_A^2
- \left( \frac{1798}{81}+\frac{56}{3}\zeta_3 \right) C_A T_R N_L
\nonumber \\ && ~~~~~~
- \left( \frac{55}{3}-16\zeta_3 \right) C_F T_R N_L
+ \frac{400}{81} T_R^2 N_L^2 .
\eea
In eq.~(\ref{coulombpot}), $\mu'=\mu e^{\gamma_E}$, where 
$\gamma_E = 0.5772...$ denotes the Euler constant.

\section{Derivation of Top Momentum Distribution}
\label{appextra}

According to the NRQCD formalism, the NNLO $\gamma t\bar{t}$
vertex in the threshold region is given by
\bea
&&
\Gamma^i(p,E) = \gamma^i \times 
\left[ C(r_0) + \frac{\Delta_{r_0}}{6m_t^2c^2} \right]
\biggl( \frac{\vc{p}^2}{m_t} - \frac{\vc{p}^4}{4 m_t^3 c^2} - \omega 
\biggr) \,
\tilde{G}_{\rm NR}(p;r_0) ,
\label{ttphotonvtx}
\\ &&
\omega = E+i\Gamma_t ,
~~~~~~~~~
E = \sqrt{s}-2 m_t c^2 .
\eea
The NRQCD Green function is defined by
\bea
&&
[ H_{\rm NR} - \omega  ] \,
G_{\rm NR}(\vc{r},\vc{r}') =
\delta(\vc{r}-\vc{r}') ,
\\
&& H_{\rm NR} = \frac{\vc{p}^2}{m_t} - \frac{\vc{p}^4}{4 m_t^3 c^2}
+ V_{\rm C}(r) + \frac{11\pi C_F a_s}{3 m_t^2 c^2} \,
\delta(\vc{r})
- \frac{C_F a_s}{2 m_t^2 c^2} \left\{ \frac{1}{r} , \vc{p}^2 \right\}
- \frac{C_F C_A a_s^2}{2m_tc^2r^2} ,
\\
&&
\tilde{G}(p;r_0) = \int d^3 \vc{r} \, e^{i \vc{p}\cdot \vc{r}}
\, G_{\rm NR}(r, r_0) ,
\eea
where $G_{\rm NR}(r, r')$ denotes the $S$-wave component of
$G_{\rm NR}(\vc{r},\vc{r}')$.
In these formulas we restored the speed of light, $c$, and
defined $a_s \equiv \alpha_s(\mu) \, c$.
Then one can identify the NLO and NNLO corrections with the
coefficients of $1/c$ and $1/c^2$, respectively, in
the series expansion of $\Gamma^i(p,E)$ in $1/c$ \cite{gr}.
The vertex renormalization constant $C(r_0)$ is determined by
matching (\ref{ttphotonvtx}) to the 2-loop $\gamma t\bar{t}$ 
on-shell vertex \cite{mc}.

From the relation \cite{mye}
\bea
&&
H_{\rm NR} =  \frac{\vc{p}^2}{m_t} + V_{\rm C}(r) 
- \frac{H_0^2}{4 m_t c^2} 
- \frac{3 C_F a_s}{4 m_t c^2} \left\{ H_0, \frac{1}{r} \right\}
+ \frac{11 C_F a_s}{12 m_t c^2} \left[ H_0, i p_r \right]
- \frac{C_F (3C_A+2C_F)a_s^2}{6 m_t c^2 r^2} ,
\nonumber \\
\\ &&
H_0 = \frac{\vc{p}^2}{m_t} - C_F \frac{a_s}{r} ,
\eea
one may find an approximate expression for the Green function
\bea
G_{\rm NR}(r, r') &\simeq&
\left[ 
1 + \frac{\omega}{2m_t c^2} 
+ \frac{3 C_F a_s}{4 m_t c^2} \biggl( \frac{1}{r} + \frac{1}{r'} \biggr)
- \frac{11 C_F a_s}{12 m_t c^2} 
\biggl( \frac{1}{r} \, \frac{d}{dr} \, r +
\frac{1}{r'} \, \frac{d}{dr'} \, r' \biggr)
\right]
G(r, r')
\nonumber \\ && 
+ \frac{1}{4 m_t c^2} \, \frac{1}{4\pi r r'} \, \delta (r-r') ,
\label{approxform}
\eea
where $G(r, r')$ is defined from a simplified Hamiltonian
in Eq.~(\ref{green}).
Using standard perturbative expansion in quantum mechanics,
one can show that
both sides of (\ref{approxform}) coincide
up to (and including) ${\cal O}(1/c^2)$ in the series expansion
in $1/c$, and that also in the vicinity
of each resonance pole, the pole position
and the residue coincide up to the same order.
One may then express the Fourier transform of (\ref{approxform})
in terms of the momentum-space Green functions defined in
eqs.~(\ref{gp1})-(\ref{gp3}).
In addition,
in the limit $r_0 \to 0$ 
one can justify a replacement
\bea
\frac{d}{dr_0} \, r_0 \, \tilde{G}(p;r_0) \to 
\left( 1 - \frac{1}{2} C_F m_t a_s r_0 \right) 
\tilde{G}(p;r_0) .
\eea 

By including the $\gamma t\bar{t}$ vertex in the
Born diagram for $e^+ e^- \to t\bar{t} \to bW^+ \bar{b}W^-$
and integrating over the $bW$ phase space, one obtains the
momentum distribution formula (\ref{momdist}).
All $r_0$-dependent factors multiplying $\tilde{G}(p;r_0)$ 
are combined
with $C(r_0)$ and included in the vertex renormalization constant
given in (\ref{momdist}).

\section{Proof of Unitarity Relation}
\label{appb}

In order to prove the unitarity relation between 
eqs.~(\ref{totcs}) and (\ref{momdist}), it is sufficient to show 
\bea
&&
\mbox{Im} \, \left[
\left( 1 + \frac{E+i\Gamma_t}{6m_t} \right)
G(r_0,r_0)
\right] 
\nonumber \\ && 
= \int {\textstyle \frac{d^3\vc{p}}{(2\pi)^3} } \, \Gamma_t \,
\left\{
\left( 1 + \frac{2E}{3m_t} \right) | \tilde{G}(p;r_0) |^2 
+ \frac{3}{2} C_F \alpha_s(\mu)^2 \,
\mbox{Re} \left[
\tilde{G}_{1/r}(p;r_0) \, \tilde{G}(p;r_0)^*
\right]
\right.
\nonumber
\\ && ~~~~~~~~~~ \left.
-\frac{11}{6}C_F \alpha_s(\mu)^2  \,
\mbox{Re} \left[
\tilde{G}_{ip_r}(p;r_0) \, \tilde{G}(p;r_0)^*
\right]
+ \frac{1}{6m_t} \, \frac{\sin (pr_0)}{pr_0} \, 
\mbox{Re} \left[
\tilde{G}(p;r_0) 
\right] \right\} .
\eea
This equality follows readily from a combination of the identities
\bea
&&
\int {\textstyle \frac{d^3\vc{p}}{(2\pi)^3} } \, \Gamma_t \,
\left\{
\left( 1 + \frac{E}{2m_t} \right) | \tilde{G}(p;r_0) |^2 
+ \frac{3}{2} C_F \alpha_s(\mu)^2 \,
\mbox{Re} \left[
\tilde{G}_{1/r}(p;r_0) \, \tilde{G}(p;r_0)^*
\right]
\right\} 
= \mbox{Im} \, G(r_0,r_0) ,
\nonumber\\
\label{id1}
\\ &&
\int {\textstyle \frac{d^3\vc{p}}{(2\pi)^3} } \, \Gamma_t \,
\mbox{Re} \left[
\tilde{G}_{ip_r}(p;r_0) \, \tilde{G}(p;r_0)^*
\right]
=0 ,
\label{id2}
\\ &&
\int {\textstyle \frac{d^3\vc{p}}{(2\pi)^3} } \, \Gamma_t \,
\frac{\sin (pr_0)}{pr_0} \, 
\mbox{Re} \left[
\tilde{G}(p;r_0) 
\right] 
= \mbox{Im} \left[ i \Gamma_t G(r_0,r_0) \right] ,
\label{id3}
\eea
and neglecting terms suppressed by ${\cal O}(\alpha_s^4)$.

\subsection*{Proof of eq.~(\ref{id1})}

Let us define an operator
\bea
G = \left[
\frac{\vc{p}^2}{m_t} + V(r) 
- \left( \omega + \frac{\omega^2}{4m_t} \right)
\right]^{-1} .
\eea
Then
\bea
{\rm Im} \, G &=& 
G^\dagger \cdot 
\frac{(G^{-1})^\dagger - G^{-1}}{2i} \cdot G
= - \, G^\dagger \cdot {\rm Im} \left[ G^{-1} \right] \cdot G
\nonumber \\
&=& G^\dagger \cdot 
\left( \Gamma_t + \frac{E \Gamma_t}{2m_t} 
+ \frac{3C_F\alpha_s}{2m_t r}
\right) \cdot G ,
\eea
where the imaginary part of any operator $X$ is defined as
${\rm Im} X = (X-X^\dagger)/(2i)$.
Sandwiching both sides by $\bra{r_0}$ and $\ket{r_0}$, and
inserting a completeness relation on the right-hand-side, one
obtains eq.~(\ref{id1}).

\subsection*{Proof of eq.~(\ref{id2})}

\bea
&&
\int {\textstyle \frac{d^3\vc{p}}{(2\pi)^3} } \, \Gamma_t \,
\mbox{Re} \left[
\tilde{G}_{ip_r}(p;r_0) \, \tilde{G}(p;r_0)^*
\right]
\nonumber \\ &&
= \int {\textstyle \frac{d^3\vc{p}}{(2\pi)^3} } \, 
\frac{\Gamma_t}{\alpha_s m_t} \,
\left[
\bra{r_0} G^\dagger \ket{p} \bra{p} ip_r \cdot G \ket{r_0}
+ \bra{r_0} G^\dagger (ip_r)^\dagger \ket{p} \bra{p} G \ket{r_0}
\right]
\nonumber \\ &&
= \frac{\Gamma_t}{\alpha_s m_t} \,
\bra{r_0} G^\dagger \, ip_r \, G + G^\dagger \, (ip_r)^\dagger \, G
\ket{r_0}
= 0 ,
\eea
where we used hermiticity of $p_r$ in the last line.

\subsection*{Proof of eq.~(\ref{id3})}

\bea
&&
\mbox{Im} \left[ i \Gamma_t G(r_0,r_0) \right]
=
\frac{\Gamma_t}{2} \, 
\bra{r_0} G + G^\dagger \ket{r_0}
\nonumber \\ &&
= \frac{\Gamma_t}{2} \, 
\int {\textstyle \frac{d^3\vc{p}}{(2\pi)^3} } \, 
\left[
\bra{r_0} G \ket{p}\braket{p}{r_0}+ 
\braket{r_0}{p}\bra{p} G^\dagger \ket{r_0}
\right]
\nonumber \\ &&
= \int {\textstyle \frac{d^3\vc{p}}{(2\pi)^3} } \, \Gamma_t \,
\frac{\sin (pr_0)}{pr_0} \, 
\mbox{Re} \left[
\tilde{G}(p;r_0) 
\right] .
\eea
Note that the $S$-wave component of $e^{i\vc{p}\cdot \vc{r}}$ is
given by $\sin (pr)/(pr)$.

\end{appendix}

\newpage

\def\app#1#2#3{{\it Acta~Phys.~Polonica~}{\bf B #1} (#2) #3}
\def\apa#1#2#3{{\it Acta Physica Austriaca~}{\bf#1} (#2) #3}
\def\npb#1#2#3{{\it Nucl.~Phys.~}{\bf B #1} (#2) #3}
\def\plb#1#2#3{{\it Phys.~Lett.~}{\bf B #1} (#2) #3}
\def\prd#1#2#3{{\it Phys.~Rev.~}{\bf D #1} (#2) #3}
\def\pR#1#2#3{{\it Phys.~Rev.~}{\bf #1} (#2) #3}
\def\prl#1#2#3{{\it Phys.~Rev.~Lett.~}{\bf #1} (#2) #3}
\def\sovnp#1#2#3{{\it Sov.~J.~Nucl.~Phys.~}{\bf #1} (#2) #3}
\def\yadfiz#1#2#3{{\it Yad.~Fiz.~}{\bf #1} (#2) #3}
\def\jetp#1#2#3{{\it JETP~Lett.~}{\bf #1} (#2) #3}
\def\zpc#1#2#3{{\it Z.~Phys.~}{\bf C #1} (#2) #3}

\newpage

\end{document}